\documentclass[12pt]{article}
\usepackage{amsfonts}
\usepackage{amssymb}
\usepackage{graphics,amsmath}

\def\hybrid{\topmargin -20pt    \oddsidemargin 0pt
        \headheight 0pt \headsep 0pt
        \textwidth 6.35in       
        \textheight 9.25in       
        \marginparwidth .875in
        \parskip 5pt plus 1pt   \jot = 1.5ex}

\hybrid

\def\baselinestretch{1.2}

\catcode`\@=11

\def\marginnote#1{}
\newcount\hour
\newcount\minute
\newtoks\amorpm
\hour=\time\divide\hour by60
\minute=\time{\multiply\hour by60 \global\advance\minute by-\hour}
\edef\standardtime{{\ifnum\hour<12 \global\amorpm={am}%
        \else\global\amorpm={pm}\advance\hour by-12 \fi
        \ifnum\hour=0 \hour=12 \fi
        \number\hour:\ifnum\minute<10 0\fi\number\minute\the\amorpm}}
\edef\militarytime{\number\hour:\ifnum\minute<10 0\fi\number\minute}

\def\draftlabel#1{{\@bsphack\if@filesw {\let\thepage\relax
   \xdef\@gtempa{\write\@auxout{\string
      \newlabel{#1}{{\@currentlabel}{\thepage}}}}}\@gtempa
   \if@nobreak \ifvmode\nobreak\fi\fi\fi\@esphack}
        \gdef\@eqnlabel{#1}}
\def\@eqnlabel{}
\def\@vacuum{}
\def\draftmarginnote#1{\marginpar{\raggedright\scriptsize\tt#1}}

\def\draft{\oddsidemargin -.5truein
        \def\@oddfoot{\sl preliminary draft \hfil
        \rm\thepage\hfil\sl\today\quad\militarytime}
        \let\@evenfoot\@oddfoot \overfullrule 3pt
        \let\label=\draftlabel
        \let\marginnote=\draftmarginnote
   \def\@eqnnum{(\theequation)\rlap{\kern\marginparsep\tt\@eqnlabel}%
\global\let\@eqnlabel\@vacuum}  }

\def\preprint{\twocolumn\sloppy\flushbottom\parindent 2em
        \leftmargini 2em\leftmarginv .5em\leftmarginvi .5em
        \oddsidemargin -.5in    \evensidemargin -.5in
        \columnsep .4in \footheight 0pt
        \textwidth 10.in        \topmargin  -.4in
        \headheight 12pt \topskip .4in
        \textheight 6.9in \footskip 0pt
        \def\@oddhead{\thepage\hfil\addtocounter{page}{1}\thepage}
        \let\@evenhead\@oddhead \def\@oddfoot{} \def\@evenfoot{} }

\def\numberbysection{\@addtoreset{equation}{section}
        \def\theequation{\thesection.\arabic{equation}}}

\def\underline#1{\relax\ifmmode\@@underline#1\else
        $\@@underline{\hbox{#1}}$\relax\fi}

\def\titlepage{\@restonecolfalse\if@twocolumn\@restonecoltrue\onecolumn
     \else \newpage \fi \thispagestyle{empty}\c@page\z@
        \def\thefootnote{\fnsymbol{footnote}} }

\def\endtitlepage{\if@restonecol\twocolumn \else \newpage \fi
        \def\thefootnote{\arabic{footnote}}
        \setcounter{footnote}{0}}  

\catcode`@=12
\relax

\def\figcap{\section*{Figure Captions\markboth
        {FIGURECAPTIONS}{FIGURECAPTIONS}}\list
        {Figure \arabic{enumi}:\hfill}{\settowidth\labelwidth{Figure
999:}
        \leftmargin\labelwidth
        \advance\leftmargin\labelsep\usecounter{enumi}}}
 \relax
\def\tablecap{\section*{Table Captions\markboth
        {TABLECAPTIONS}{TABLECAPTIONS}}\list
        {Table \arabic{enumi}:\hfill}{\settowidth\labelwidth{Table
999:}
        \leftmargin\labelwidth
        \advance\leftmargin\labelsep\usecounter{enumi}}}
 \relax
\def\reflist{\section*{References\markboth
        {REFLIST}{REFLIST}}\list
        {[\arabic{enumi}]\hfill}{\settowidth\labelwidth{[999]}
        \leftmargin\labelwidth
        \advance\leftmargin\labelsep\usecounter{enumi}}}
 \relax
\makeatletter
\newcounter{pubctr}
\def\publist{\@ifnextchar[{\@publist}{\@@publist}}
\def\@publist[#1]{\list
        {[\arabic{pubctr}]\hfill}{\settowidth\labelwidth{[999]}
        \leftmargin\labelwidth
        \advance\leftmargin\labelsep
        \@nmbrlisttrue\def\@listctr{pubctr}
        \setcounter{pubctr}{#1}\addtocounter{pubctr}{-1}}}
\def\@@publist{\list
        {[\arabic{pubctr}]\hfill}{\settowidth\labelwidth{[999]}
        \leftmargin\labelwidth
        \advance\leftmargin\labelsep
        \@nmbrlisttrue\def\@listctr{pubctr}}}
 \relax
\makeatother
\newskip\humongous \humongous=0pt plus 1000pt minus 1000pt

\newif\ifdtup

\relax

\def\be{\begin{equation}}
\def\ee{\end{equation}}
\def\ba{\begin{eqnarray}}
\def\ea{\end{eqnarray}}

\def\no{\noindent}

\def\IR{\relax{\rm I\kern-.18em R}}
\def\II{\relax{\rm 1\kern-.35em1}}

\renewcommand{\theequation}{\thesection.\arabic{equation}}
\csname @addtoreset\endcsname{equation}{section}

\def\IR{\relax{\rm I\kern-.18em R}}
\def\inv{^{\raise.15ex\hbox{${\scriptscriptstyle -}$}\kern-.05em 1}}

\def\dif{{\textnormal d}}

\begin{document}

\begin{titlepage}
\begin{center}

\vskip .5in

{\LARGE Minimal surfaces with mixed three-form flux}
\vskip 0.4in

{\bf Rafael Hern\'andez},  \phantom{x} {\bf Juan Miguel Nieto} \phantom{x} and \phantom{x} {\bf Roberto Ruiz} 
\vskip 0.1in

Departamento de F\'{\i}sica Te\'orica \\
Universidad Complutense de Madrid \\
$28040$ Madrid, Spain \\
{\footnotesize{\tt rafael.hernandez@fis.ucm.es, juanieto@ucm.es, roruiz@ucm.es}}

\end{center}

\vskip .4in

\centerline{\bf Abstract}
\vskip .1in
\no
\noindent
We study minimal area world sheets ending on two concentric circumferences on the boundary of Euclidean $AdS_{3}$ with mixed Ramond-Ramond and Neveu-Schwarz-Neveu-Schwarz three-form fluxes. 
We solve the problem by reducing the system to a one-dimensional integrable model. We find that the Neveu-Schwarz-Neveu-Schwarz flux term either brings the surface near to the boundary 
or separates the circumferences. In the limit of pure Neveu-Schwarz-Neveu-Schwarz flux, the solution adheres to the boundary in the former case and the outer radius diverges in the latter. 
We further construct the underlying elliptic spectral curve, which allows us to analyze the deformation of other related minimal surfaces. 
We show that in the regime of pure Neveu-Schwarz-Neveu-Schwarz flux the elliptic curve degenerates.

\vskip .4in
\noindent

\end{titlepage}

\vfill
\eject

\def\baselinestretch{1.2}


\baselineskip 20pt


\section{Introduction}

The AdS/CFT correspondence states that the strong coupling limit of the expectation value of Wilson loop observables can be described by classical string solutions. 
In particular, it is given by the area of the minimal surface swept by a string subtending the path determined by the loop on the boundary of the anti-de Sitter space~\cite{Maldacena}. 
A general class of minimal surfaces was found in reference~\cite{DF} using a periodic ansatz that allowed the authors to reduce the analysis of the nonlinear sigma-model for the string to the construction 
of solutions to one-dimensional integrable systems. This periodic ansatz is closely related to the ansatz introduced in~\cite{NR} to transform the analysis of the energy spectrum 
of circular strings spinning in $AdS_{5} \times S^{5}$ into the study of periodic solutions to the Neumann-Rosochatius integrable system, which is a 
model of oscillators on a sphere. In reference~\cite{HN} it was shown that the analysis of closed strings spinning in $AdS_{3} \times S^{3} \times T^{4}$ with a mixture 
of Ramond-Ramond (R-R) and Neveu-Schwarz-Neveu-Schwarz (NS-NS) three-form fluxes can be performed by means of an integrable deformation of the Neumann-Rosochatius mechanical system (see~\cite{CZ}-\cite{BBP} 
for other solutions and various problems regarding the integrability of the AdS$_3$/CFT$_2$ correspondence with mixed fluxes). The purpose of the present article is to extend 
the approach of~\cite{DF} to find minimal area surfaces ending on two concentric circumferences in the presence of NS-NS flux using the integrable deformation 
of their associated mechanical system. 

Besides the study of some aspects of the AdS/CFT correspondence in terms of simple mechanical models, the integrability of the problem has also lead 
to several complementary series of developments. One of them comes from the use of algebraic spectral curves to encode the dynamics of spinning string solutions, 
as was first noted in~\cite{Kazakov}. The construction of the spectral curves starts from the existence of nontrivial holonomies on the world sheet of the closed string. 
Therefore, when trying to construct algebraic curves for or world sheets ending on loops one should face the problem that the holonomies may be trivial therein. 
However, factorizability of some classes of solutions allows one to circumvent the absence of non-trivial holonomies since the construction just requires the local form 
of the Lax connection whenever it is present~\cite{Janik,DekelII}. As our minimal solution with mixed flux preserves factorizability, we will follow this path 
here and adapt the procedure applied in~\cite{Dekel} to the case of a nonvanishing NS-NS three-form.

The plan of the remaining part of the paper is as follows. In section 2 we will employ the integrability of the problem to construct minimal area surfaces ending on an annulus
with mixed R-R and NS-NS fluxes. We will show that the presence of the NS-NS flux term either reduces the distance that the string is allowed to enter into the bulk or increases 
the separation between the circumferences on the boundary. In the limit of pure NS-NS flux, the action reduces to a supersymmetric Wess-Zumino-Witten model. 
We will prove that in this limit the world sheet is stuck on the boundary in the former case and that the radius of the outer circumference diverges in the latter. 
We will also discuss the existence of D1-string solutions related to these surfaces. In section 3 we will tackle the problem 
by constructing the associated spectral curve. We will use this result to recover the spectral curves of other configurations in some particular limits. We will find the deformation of the underlying elliptic surface 
by the flux term and show that either one or both cycles of the complex torus degenerate when the R-R flux vanishes. In section 4 we summarize our results and comment on some open issues 
and future perspectives. 


\section{Minimal area surfaces}

In this section we will construct the classical string solutions corresponding to a minimal surface that ends on two concentric circumferences on the boundary of the anti-de Sitter space in the presence of flux. 
In particular, we will be interested in classical string solutions embedded in Euclidean $AdS_{3}$ with nonvanishing NS-NS three-form flux. We will write the background metric as
\be
\dif s^2 = z^{-2} ( \dif r^2 + r^2\dif \theta^2 + dz^2) \ , 
\label{Polar metric}
\ee
and the NS-NS B-field as 
\be
B = i q z^{-2} \, r\dif r\wedge \dif \theta \ , 
\label{Polar B-field}
\ee
where $|q| \leq 1$. The case where $q = 0$ corresponds to the limit of pure R-R flux, while we are left with pure NS-NS flux when $|q|=1$ is set. 
In the above coordinates, the surface ending on the annulus is described by the ansatz~\cite{DF}~\footnote{Here and throughout this article, $\tau$ denotes the Euclidean world sheet time coordinate. 
}
\be
r(\tau,\sigma)= r(\tau) \ , \quad \theta(\tau,\sigma)=k\sigma \ , \quad z(\tau,\sigma)=z(\tau) \ ,
\label{Ansatz}
\ee
together with the boundary conditions
\be
r(\tau_{i},\sigma)= R_{i} \ , \quad \theta(\tau_{i},\sigma)=k\sigma \ , \quad z(\tau_{i},\sigma)=0 \ , \quad i=1,2 \ ,
\label{Boundary}
\ee
with $R_{i}$ the radii of the circles, and $k$ the winding index along the circumferences. When we enter this ansatz into the Euclidean world sheet action 
in the conformal gauge, we find a one-dimensional integrable mechanical model with Lagrangian
\be
L = z^{-2} ( \dot{r}^2 +\dot{z}^2  - 2qkr\dot{r} + r^2 k^2) \ .
\label{S}
\ee
The integrability of this model follows from the existence of a first integral of motion associated to the dilatation invariance of the action. 
This integral parametrizes the solution and reads 
\be
p = z^{-2} ( r\dot{r} +z\dot{z} -q k r^2 ) \ .
\label{Equation}
\ee
In order to solve this equation, we will also need the vanishing-energy constraint imposed by the conformal gauge condition, 
\be
\dot{r}^2 + \dot{z}^2 - k^2 r^2 = 0 \ .
\label{Constraint}
\ee
For convenience, we will introduce new variables $u$ and $v$, defined through
\be
z = \frac{u e^v}{\sqrt{1+u^2}} \ , \quad r=\frac{e^v}{\sqrt{1+u^2}} \ .
\label{Change}
\ee
In these coordinates, equations (\ref{Equation}) and (\ref{Constraint}) become
\ba
\dot{u}^2 & \!\! = \!\! & (1-q^2)k^2+(k^2-2qkp)u^2-p^2u^4 \ , \label{Polynomial} \\
\dot{v} & \!\! = \!\! & \frac{u^2p+qk}{1+u^2} \ , \label{vdot}
\label{Radii}
\ea
while the boundary conditions read
\be
u(\tau_{i}) = 0 \ , \quad v(\tau_{i})=\log R_{i} \ , \quad i=1,2 \ .
\ee
The problem is therefore reduced to finding the solution to the differential equation~(\ref{Polynomial}). This equation defines a genus-one complex surface, 
and thus the solution can be written in terms of elliptic functions. 

In order to find the solutions to (\ref{Polynomial}), we will separate the cases with $\left|p\right|>0$ and $p=0$. We will consider first the case with $\left|p\right|>0$. 
Equation~(\ref{Polynomial}) can be then rewritten as
\be
\label{Solve}
\dot{u}^2=p^2(u_{-}^2+u^2)(u_{+}^2-u^2)=P_{4}(u) \ ,
\ee
with squared roots
\be
u_{\pm}^2 = \frac{ \pm(k^2-2qkp) + \sqrt{(k^2-2qkp)^2+4(1-q^2)k^2p^2}}{2p^2} \ .
\ee
These squared roots satisfy $u_{\pm}^{2} \geq 0$. In the limit where $q = \pm 1$, the equality is reached by $u_{+}^{2}$ if we choose $k^2 \mp 2kp \leq 0$ or by $u_{-}^{2}$ if $k^2 \mp 2kp \geq 0$. 
Since $q = \pm 1$ corresponds to the situation where two of the roots in the polynomial $P_{4}(u)$ coalesce, one of the cycles of the complex surface degenerates in this limit. 

The coordinate $u$ varies from the boundary at $u=0$ to a maximum value $u_{+}$ corresponding to the turning point where $\dot{u}=0$. This is the branch of the solution 
with $\dot{u}>0$. Beyond the turning point, we have to move to the branch that extends from $u_{+}$ to the boundary, where $\dot{u}<0$.
We choose the interval of integration to go from $\tau_{1}=0$ to $\tau_{2}=T$ so that the turning point is located at $\tau=T/2$. If we define
\be
\alpha(u)\equiv\textnormal{arcsin}_{\left[0,\frac{\pi}{2}\right]}\left(\sqrt{\frac{u_{-}^2+u_{+}^2}{u_{-}^2+u^2}}\frac{u}{u_{+}}\right) \ , \quad \kappa\equiv\frac{u_{+}}{\sqrt{u_{-}^2+u_{+}^2}} \ ,
\ee
the solution to equation (\ref{Solve}) can be expressed as
\ba
\tau & \!\! = \!\! & \frac{F (\alpha , \kappa)}{{\left|p\right|}\sqrt{u_{-}^2+u_{+}^2}} \ , \quad \textnormal{when} \:\: 0\le\tau\le \frac{T}{2} \ , \label{tau1} \\
\tau & \!\! = \!\! & \frac{2K(\kappa ) - F ( \alpha , \kappa )}{{\left|p\right|}\sqrt{u_{-}^2+u_{+}^2}} \ , \quad \textnormal{when} \: \: \frac{T}{2}\le\tau\le T \label{tau2} \ ,
\ea
where $F(\alpha,\kappa)$ and $K(\kappa)$ are, respectively, the incomplete and complete elliptic integrals of the first kind. 
Therefore, the length of the interval of integration is given by
\be
T = \frac{2K(\kappa)}{{\left|p\right|}\sqrt{u_{-}^2+u_{+}^2}} \ .
\label{Length}
\ee
Inverting relations (\ref{tau1}) and (\ref{tau2}), we conclude that 
\be
\label{Stuck}
u^2(\tau)=\frac{u_{-}^2u_{+}^2}{u_{-}^2+u_{+}^2}\textnormal{sd}^2 \big( \left|p\right|\sqrt{u_{-}^2+u_{+}^2}\tau , \kappa \big) \ , \quad \textnormal{with} \:\: 0\le\tau\le T \ ,
\ee
where we have used the parity and semiperiodicity properties of the Jacobi function 
\be
\textnormal{sd}(-z,\kappa)=-\textnormal{sd}(z,\kappa) \ , \quad \textnormal{sd}(z+2K(\kappa),\kappa)=-\textnormal{sd}(z,\kappa) \ .
\ee
We can now integrate equation (\ref{vdot}) to find 
\be
\begin{split}
\label{Second expression}
v (\tau) = \frac{u^2_{-}(p-qk)\Pi\left(\textnormal{am}\left(\left|p\right|\sqrt{u_{-}^2+u_{+}^2}\tau,\kappa\right),n,\kappa\right)}{{\left|p\right|}(1-u^2_{-})\sqrt{u_{-}^2+u_{+}^2}}
-\frac{u^2_{-}p-qk}{1-u^2_{-}}\tau+\log R_{1} \ , 
\end{split}
\ee
where $\Pi(\alpha,n,\kappa)$ is the incomplete elliptic integral of the third kind, with characteristic 
\be
n = \frac{u^2_{+}(1-u_{-}^2)}{u_{-}^2+u_{+}^2} \ .
\ee
The ratio of the radii $R=R_{2}/R_{1}$ can be obtained by setting $\tau=T$ in (\ref{Second expression}),
\be
\label{Exponential}
R = \exp \left(\frac{2u^2_{-}(p-qk)\Pi(n,\kappa)-2\left(u^2_{-}p-qk\right)K(\kappa)}{{\left|p\right|}(1-u^2_{-})\sqrt{u_{-}^2+u_{+}^2}}\right) \ ,
\ee
where we have used (\ref{Length}) and $\Pi(n,\kappa)$ is the complete elliptic integral of the third kind.

We will consider now the case of pure NS-NS flux, where $q=1$ (the limit where $q=-1$ can be obtained from the $q=1$ case by performing the substitution $k\rightarrow-k$). 
We must first note that the solution (\ref{Stuck}) displays sharply different behaviors depending on the sign of $k^2-2kp$. In the region where $k^2-2kp<0$, we find
\be
u_{+}^2 \rightarrow 0 \ , \quad u_{-}^2 \rightarrow - \frac{k^2-2kp}{p^2} \ .
\ee
According to (\ref{Stuck}), the world sheet penetrates into the bulk from the boundary at $u=0$ until it reaches $u=u_{+}$. 
As the flux grows, the position of the turning point $u_{+}$ decreases. In the limit of pure NS-NS flux, the world sheet adheres entirely to the boundary. 
Moreover, both the elliptic modulus $\kappa$ and the elliptic characteristic $n$ vanish. Therefore, it is straightforward to show that the length of the interval 
$T$ tends to~\footnote{Subsequently, barred quantities are evaluated in the limit of pure NS-NS flux.}
\be
\bar{T}=\frac{\pi}{\sqrt{\left|k^2-2kp\right|}} \ ,
\ee
and the ratio $R$ of the radii tends to
\be
\bar{R} = \exp\left(\frac{k\pi}{\sqrt{\left|k^2-2kp\right|}}\right) = \exp \left(k\bar{T}\right) \ .
\ee
On the contrary, if $k^2-2kp>0$ we find
\be
u_{+}^2 \rightarrow \frac{k^2-2kp}{p^2} \ , \quad u_{-}^2\rightarrow 0 \ .
\ee
In this case the elliptic modulus $\kappa$ tends to one, and from~(\ref{Length}), we thus conclude that the integration length goes to infinity. 
In order to understand in more detail this case, it is convenient to shift the world sheet time coordinate through 
\be
\tau\rightarrow\tau+\frac{K(\kappa)}{\left|p\right|\sqrt{u_{-}^2+u_{+}^2}} \ ,
\ee 
which corresponds to choosing its range from $-T/2$ to $T/2$, before applying the limit to equation (\ref{Stuck}). After performing the shift, we find
\be
u^2(\tau)=u^2_{+}\textnormal{cn}^2 \big( \left|p\right|\sqrt{u_{-}^2+u_{+}^2}\,\tau,\kappa \big) \ , \quad \hbox{with} \:\: - \frac{T}{2}\le\tau\le \frac{T}{2} \ ,
\ee
and thus
\be
u^2(\tau)\rightarrow u^2_{+}\textnormal{sech}^2 \big( \sqrt{k^2-2kp}\,\tau \big) \ , \quad \hbox{with} \:\: - \infty<\tau<\infty \  .
\ee
From equation (\ref{Exponential}), we conclude that $R$ diverges if $k$ is positive and that it goes to zero if $k$ is negative. In fact, asymptotically, it is satisfied, 
\be
\bar{R} \simeq \exp\left(k\bar{T}\right) \ ,
\ee
as in the case where $k^2-2kp\le0$. Finally, in the threshold case where $k^2-2kp=0$, we find that $u_{\pm}^2\rightarrow 0$, 
which corresponds to a world sheet stuck on the boundary that subtends an annulus for which the ratio of the radii either diverges or tends to zero.
 
We will now move to the case where $p=0$, which is a particular limit of the general solution presented above. Accordingly, the elliptic surface associated to (\ref{Solve}) reduces its genus by one. 
Moreover, it follows from equation (\ref{Exponential}) that in the limit of pure R-R flux the two radii coincide, and we are led to a circle on the boundary. Either applying the limit to (\ref{Stuck}) 
and (\ref{Second expression}) or starting from (\ref{Polynomial}) and (\ref{Radii}) and substituting $p=0$ therein, we find that
\begin{align}
\label{Third}
&u(\tau)=\sqrt{1-q^2}\sinh\left(\left|k\right|\tau\right) \ ,\\
\label{Fourth}
&v(\tau)=\textnormal{arctanh}\left(q\tanh\left(k\tau\right)\right)+\log R_{1} \ .
\end{align}
If $q=\pm1$, the world sheet adheres to the boundary, and the radius of the external circumference diverges. This fact is consistent with the previous discussion on the pure NS-NS flux limit, since $k^2>0$ is always satisfied.

We must note that equations (\ref{Radii}) and (\ref{Solve}) with the boundary conditions (\ref{Boundary}) admit an infinite class of solutions that corresponds to a world sheet stuck on the boundary, 
and which are absent everywhere but in the pure NS-NS flux limit. Integrating (\ref{Radii}) with $u=0$ leads to $v=k\tau$. It follows then that the ratio of the radii and the length of the interval 
are related through $\bar{R}=\exp{\left(k\bar{T}\right)}$. Given the winding index $k$, any value of $\bar{R}$ can be achieved for these solutions by choosing properly the length $\bar{T}$ and vice versa.

We will conclude this section with a brief discussion on the limiting cases of pure R-R and NS-NS flux 
by means of the $SL(2,\mathbb{Z})$ symmetry of type IIB strings \cite{Obers}. 
Let us first note that since $AdS_{3}\times S^{3} \times T^{4}$ supported by mixed flux emerges as the near horizon limit of a brane array~\cite{Giveon}, 
it does not need to respect the full $SL(2, \mathbb{Z})$ symmetry. Therefore, we have to limit ourselves to those transformations that do not modify its field content. 
In particular, given that both the dilaton and the R-R scalar vanish initially~\cite{Hoare}, they must also vanish after performing the $SL(2, \mathbb{Z})$ transformation. If we take into account 
the properties of the fields, we realize that just trivial and S-duality transformations are allowed~\cite{Kluson}. Since S-duality interchanges F1- and D1-string numbers, we conclude 
that the F1-strings that we have found with vanishing R-R flux imply the existence of D1-strings with vanishing NS-NS flux which display an behavior analogous to the former and vice versa. 
We may also draw this conclusion starting from the Dirac-Born-Infeld action with a Wess-Zumino term for the R-R two-form. The Euler-Lagrange equations for the gauge potential 
imply that the world sheet Abelian electric field $\mathcal{F} = B + 2 \pi\alpha'F$ is related to a constant field and that it vanishes when the latter does~\cite{Kluson}. Furthermore, 
the boundary conditions on the spatial world sheet coordinate imply that the constant field is semiclassically quantized in terms of the F1-string number of the solution~\cite{Bachas}, 
which is zero for D1-strings. Since $\mathcal{F}$ vanishes, we could apply the second order formalism in the action. Therefore, the ansatz~(\ref{Ansatz}) with the periodic boundary 
conditions (\ref{Boundary}) would have led us to the Lagrangian (\ref{S}) with the NS-NS flux term replaced by the one for the R-R flux, i.e., with $q$ replaced by $\pm\sqrt{1-q^2}$. 


\section{Algebraic curves and minimal surfaces}

In this section, we will extend the analysis of the case of pure R-R flux in~\cite{Dekel} to construct the deformation by the NS-NS flux term 
of the algebraic elliptic curve that describes the minimal surfaces ending on an annulus on the boundary of Euclidean $AdS_{3}$ that we have obtained above. 
Under the presence of mixed R-R and NS-NS fluxes the Lax connection 
is deformed~\cite{CZ,Babichenko}, and therefore the spectral curve defined by the quasimomenta~\cite{Babichenko,Janik,DekelII} should also be  deformed. In order to find this deformation, 
we will start by fixing the gauge condition for the coset group element to $g_{L}\oplus g_{R}=1\oplus g$ and then performing a Wick rotation both in the space-time 
and the world sheet coordinates. We will choose the group element for Euclidean $AdS_{3}$ as 
\be
\label{Representative}
g=  S^{-1} \frac{1}{z}
\begin{pmatrix} r^{2}+z^{2} & r \\
r & 1
\end{pmatrix}
S \ , 
\ee
where $S = e^{-i (\theta/2) \sigma_{3}}$. When we take into account the ansatz (\ref{Ansatz}), the parametrization of the group element becomes 
factorizable, i.e., $g(\tau,\sigma)=S^{-1}(\sigma) g(\tau,0) S(\sigma)$. This property is inherited by the Maurer-Cartan form $j=g^{-1}\textnormal{d}g$, the components of which read
\ba
j_\tau & \!\! = \!\! & S^{-1}  \frac{1}{z^2} 
\begin{pmatrix}
 {r\dot{r}+z \dot{z}} & \dot{r} \\
 \left(z^2-{r^2}\right) \dot{r}-2 r z \dot{z} & -\left(r\dot{r}+z \dot{z} \right)
\end{pmatrix}
S \ , \label{jt} \\
j_\sigma & \!\! = \!\! &  S^{-1} \frac{ik}{z^2}
\begin{pmatrix}
  r^2 & r\\
 -\left(r^2+z^2\right)r & -r^2
\end{pmatrix} 
S \ . \label{js} 
\ea
The components of the relevant part of the deformed Lax connection are~\cite{Babichenko}
\be
A_{\tau} = \frac{(qx +\tilde{q}  )j_\tau +  ix j_\sigma}{\tilde{q} \left(s-x\right)\left(1/s+x\right)} \ , \quad 
A_{\sigma} = \frac{(qx + \tilde{q} )j_\sigma -ix j_\tau}{\tilde{q} \left(s-x\right)\left(1/s+x\right)} \ ,
\ee
where $x$ denotes the spectral parameter, and we have introduced $s=\sqrt{(1-q)/(1+q)}$ and $\tilde{q} = \sqrt{1-q^{2}}$.~\footnote{Note that integrability just implies the condition $q^2+\tilde{q}^2=1$\cite{CZ}. 
However, for concreteness, it will be assumed throughout the subsequent discussion that $0\le q,\tilde{q}\le 1$. 
}
We must stress that the representation of the classical solution in terms of a complex surface does not require the explicit form of the monodromy 
matrix because $S^{-1}\textnormal{d}S = - i(k/2)\sigma_{3}\textnormal{d}\sigma$~\cite{DekelII}. It is just the local form of the Lax connection that is implied. If we define
\be
\label{Definition}
L(\tau, x) = A_{\sigma}(\tau, \sigma=0, x) + i \frac{k}{2}\sigma_{3} \ ,
\ee
the quasimomenta $\pm p(x)$ that define the complex surface read~\cite{Dekel}
\be
\label{Quasimomenta}
p(x)=2\pi\sqrt{\textnormal{det}\, L (\tau, x)}+k\pi \ .
\ee
Using now equations (\ref{jt})-(\ref{Definition}), we immediately conclude that
\ba
\textnormal{det} \, L & \!\! = \!\! & \frac{1}{4\tilde{q}^2(s-x)^2\left(1/s+x\right)^2z^2} \Big[ \vphantom{\left(-\tilde{q}  x ^2 + \kappa +2 q x\right)^2}
4 x ^2 \left(\dot{r}^2+\dot{z}^2\right)-4 k x  \left(r\dot{r}+z\dot{z}\right) \left[\tilde{q}  \left(1-x ^2\right)+2 x  q \right]  \nonumber \\
& \!\! + \!\! & 4 k^2 x  (\tilde{q} +q x) (q-\tilde{q} x)r^2+k^2 \left[\tilde{q}\left(1-x ^2\right)+2xq\right]^2 z^2
\Big] \ ,
\ea   
that can be rewritten as
\ba
\label{ciarbeglA}            
\textnormal{det}\, L & \!\! = \!\! & \frac{k}{4\tilde{q}^2(s-x)^2\left(1/s+x\right)^2} \Big[ k\tilde{q}^2x ^4 + 4\left( p -q k
\right) \tilde{q}x ^3   \nonumber \\
& \!\! + \!\! & \left(6 q^2 k-2 k-8 q p\right)x ^2 - 4 \left( p - q k \right) \tilde{q}x + k \tilde{q}^2 \Big ] \ ,
\ea
where we have made use of relations (\ref{Equation}) and (\ref{Constraint}). We must note that the time dependence is lost as ensured by the isospectral property 
of the spectrum of the monodromy matrix. In the limit of pure NS-NS flux, the quasimomenta trivializes abruptly,
\be
\bar{p}=2\pi\sqrt{k^2-2pk}+k\pi \ ,
\ee
becoming independent of the spectral parameter. Note that the sign of the combination $k^2-2kp$, which as we have discussed in the previous section governs the behavior of the solution in the limit $q=1$, 
may be regarded as a reality condition on the shifted quasimomenta $\bar{p}-k\pi$.

The equation of the algebraic curve defined by the quasimomenta can be constructed from (\ref{ciarbeglA}) by setting aside the poles of the Lax connection~\cite{DekelII},
\be
\textnormal{det}\left[y^2-\left(s-x\right)\left(1/s+x\right) L \right]=0 \ ,
\ee
which leads to 
\be
y^2 = \frac{k^2}{4}x ^4+\frac{k\left( p -q k\right)}{\tilde{q}}x ^3+ \frac{k\left(3 q^2 k- k-4 q p\right)}{2\tilde{q}^2}x ^2 -\frac{k\left( p -q k\right)}{\tilde{q}} x + \frac{k^2}{4} \ .
\ee
In order to understand the behavior of this algebraic curve in the limits of interest we will redefine $x$ and $y$ to bring 
the previous equation into the Weierstrass form (see, for instance, reference~\cite{Chandrasekharan}),
\be
\label{Normal}
y^2 = 4x^3 - g_{2}x - g_{3} \ ,
\ee
where the functions $g_{2}$ and $g_{3}$ are given by
\ba
g_{2} & \!\!\! = \!\!\! & \frac{64}{3} \left[\left(1 - 3 q^2 + 3 q^4\right)k^4 + 2 pq \left(1 - 3 q^2 \right) k^3 +  q^2 \left(4 p^2+3 - 3 q^2\right) k^2 - 3 \tilde{q}^2 \left(q p k  - p^2\right) \right] \ , \nonumber \\
g_{3} & \!\!\! = \!\!\! & \frac{256}{27} k \left(k - 2 p q \right) \big[\left(2 - 9 q^2 + 9 q^4\right) k^4 + 2 p q \left(5 - 9 q^2 \right) k^3 + q^2 \left( 8 p^2 + 9 - 9 q^2\right) k^2 \nonumber \\
& \!\!\! - \!\!\! & 9 \tilde{q}^2 ( 2 q p k  -p^2) \big] \ . 
\ea
The $j$-invariant of the algebraic curve can then be found through
\be
j = \frac{1728 \, g_{2}^{3}}{\Delta} \ ,
\label{invariant}
\ee          
where the modular discriminant $\Delta = g_{2}^{3}-27g_{3}^{2}$ is 
\ba
\Delta & \!\! \!= \!\!\! & 27 \tilde{q}^4 \Big[ k^4 + 4 p^2 + 4 q p k (k^2-2) + 4 q^2 (1- k^2) ( \tilde{q}^2 k^2 -2 k p q -  p^2 ) \Big] \nonumber \\
& \!\!\! \times \!\!\! & \left[ p^2 +  q k (1- k^2) ( q k -2 p) \right]^2 \ .
\ea
The $j$-invariant depends on the modular parameter $\omega=\omega_{2}/\omega_{1}$, with $\omega_{1}$ and $\omega_{2}$ 
the periods of the complex torus. This function is invariant under modular transformations and 
discriminates between elliptic curves over the complex numbers 
belonging to different isomorphism classes.

In the case where $q=0$ and $k= \pm 1$, the algebraic curve (\ref{Normal}) describes other minimal surfaces in some limiting cases, namely, the world sheet subtending a circle when $p=0$ 
and two parallel lines when $p$ tends to infinity. From the point of view of the $j$-invariant, the circle corresponds to $j=\infty$, where one of the cycles 
of the torus collapses, and the  two parallel lines correspond to $j=1728$, where the group of automorphisms of the curve is enhanced. The solution corresponding to the last case 
is obtained directly from ours by making the two radii tend to infinity while keeping their radial distance finite~\cite{Dekel}. 

To obtain the extension of the circle under NS-NS flux, we will set $k=\pm 1$ and solve $1/j=0$ for $p$. 
If we impose that $p=0$ when $q=0$, we still have $p=0$ as a solution for any value of the flux. The algebraic curve then reads
\begin{equation}
y^2=4x^3-\frac{64}{3}x-\frac{512}{27} \ ,
\end{equation}
which is not deformed by the flux term. We stress, on the basis of the discussion above (\ref{Third}) and (\ref{Fourth}) and the analysis performed here, that the most natural extension 
of the circle with nonvanishing NS-NS flux involves its transformation into an annulus at the boundary. In the two parallel lines case, we can proceed analogously. 
If we set $k= \pm 1$ and solve $j=1728$ for $p$ 
requiring that $p\rightarrow \infty$ when $q=0$, we find that $p= \pm 1/2q$. The algebraic curve in the Weierstrass form now reads
\be
y^2=4x^3-16\frac{1-q^2}{q^2}x \ .
\ee
Note that the value of $p$ in the limit of pure NS-NS flux is that of the threshold $k^2-2kp=0$, with $k=\pm 1$. We should also mention that 
other solutions, with different values for $p$ and absent in the pure R-R regime and that share the same $j$-invariant with the ones considered above, 
can be obtained by allowing other values for $k$ and loosening the condition on $p$ when $q=0$.

Let us focus now on the limit of pure NS-NS flux. In this case, $g_{2}$ and $g_{3}$ become
\be
\bar{g}_{2}=\frac{64}{3}\left(k^2-2kp\right)^2 \ , \quad \bar{g}_{3}=\frac{512}{27}\left(k^2-2kp\right)^3 \ ,
\ee
and the elliptic discriminant vanishes. In terms of the roots $e_{i}$ of (\ref{Normal}), we get
\ba
\bar{e}_{1} & \!\! = \!\! & -\frac{2}{3}\left(k^2-2kp\right)+2\left|k^2-2kp\right| \ , \nonumber \\
\bar{e}_{2} & \!\! = \!\! & -\frac{2}{3}\left(k^2-2kp\right)-2\left|k^2-2kp\right| \ , \\
\bar{e}_{3} & \!\! = \!\! & -\frac{4}{3}\left(k^2-2kp\right) \ , \nonumber
\ea
and hence either $e_{1}$ or $e_{2}$ coalesces with $e_{3}$, depending on the sign of $k^2-2kp$. The superposition of two roots, or equivalently the vanishing of the elliptic discriminant, 
implies a divergent $j$-invariant, which corresponds to the modular parameter $\bar{\omega}=i\infty$. Therefore, one of the cycles of the complex torus degenerates, 
and the latter gets pinched, becoming topologically equivalent to a complex sphere with two points identified. If the condition $k^2-2kp=0$ is also satisfied, $\bar{g}_{2}=\bar{g}_{3}=0$, 
and the three roots coalesce to zero. The $j$-invariant is then $\bar{j}=1728$, which corresponds to $\bar{\omega}=i$. In the torus picture, the two points identified become the same 
when such a threshold is reached, and thus the pinched torus turns into a sphere. 


\section{Conclusions}
 
In this article, we have constructed minimal surfaces that end on two concentric circumferences in Euclidean $AdS_{3}$ with mixed R-R and NS-NS three-form fluxes. 
We have shown that the effect of the NS-NS flux term is either to decrease the distance that the minimal surface is allowed to enter into the bulk or to increase the relative radii of the circles at the boundary. 
By taking the limit of pure NS-NS flux, we have found that in the first case the string world sheet remains stuck on the boundary while in the second case the external radius diverges. 
We have also argued the existence of analogous classical D1-string solutions in the regime of pure R-R flux. We have further constructed the spectral curve that describes the solutions
that we have found, studying various limiting cases that should describe the deformation under mixed flux of other relevant world sheets. Finally, we have found that the pure NS-NS limit of the solution 
may be regarded as a degenerate limit of its associated complex torus.

The family of minimal surfaces constructed here should be straightforwardly generalizable when nonvanishing angular momentum in the sphere is included. In particular, these solutions encompass the world sheet stretching between a circle and a point in the limit of pure R-R flux when one of the circumferences collapses. As the nontrivial mixture of fluxes would again deform these world sheets, 
their study may shed light on the limit of pure NS-NS flux of minimal surfaces, indicating if the behavior of the world sheet we have encountered persists. Moreover, this extension may open 
the possibility of finding other classes of solutions in the mixed flux regime that would be otherwise absent. 

It would be also interesting to study if any phase transition occurs either for the minimal area surfaces that we have found or for their extension endowed with angular momentum in the sphere. Regarding 
the on-shell regularized action, the authors of \cite{Gross} found a phase transition between connected and disconnected world sheets in $AdS_{5} \times S^{5}$ depending on the axial distance between 
two concentric circumferences on the boundary. Performing an analogous analysis, it could be possible to establish which are the preferential solutions in terms of their action, including the 
different classes found here in the regime of pure NS-NS flux. Furthermore, it could happen that the flux mixing term might tune some possible phases of minimal surfaces.


\vspace{8mm}

\centerline{\bf Acknowledgments}

\vspace{2mm}

\no
The work of R.~H. is supported by grant FPA2014-54154-P and by BSCH-UCM through grant GR3/14-A 910770.


\end{document}